# Reversible optical data storage below the diffraction limit

Richard Monge[1,2,*], Tom Delord[1,*], and Carlos A. Meriles[1,2,†]

Color centers in wide-bandgap semiconductors feature metastable charge states that can be interconverted with the help of optical excitation at select wavelengths. The distinct fluorescence and spin properties in each of these states have been exploited to show storage of classical information in three dimensions, but the memory capacity of these platforms has been thus far limited by optical diffraction. Here, we leverage local heterogeneity in the optical transitions of color centers in diamond to demonstrate selective charge state control of individual point defects sharing the same diffraction-limited volume. Further, we apply this approach to dense color center ensembles, and show rewritable, multiplexed data storage with large areal density. These results portend alternative approaches to information processing in the form of devices with enhanced optical storage capacity.

Wide-bandgap semiconductors host impurities and other point defects whose absorption at select optical wavelengths are known to confer gems their signature hues. These "color centers" can take different charge states, each featuring characteristic optical and spin properties. Relevant examples are the nitrogen-vacancy (NV) and silicon-vacancy (SiV) centers in diamond, or the di-vacancy and the silicon-vacancy centers in silicon carbide, presently attracting broad interest for applications in quantum information processing and nanoscale sensing[1,2]. Laser illumination at select wavelengths induces ionization and recombination processes that can drive the color center into desired states of charge almost deterministically[3]. Injection of carriers produced by light-driven charge cycling of color centers serves as a tool to probe charge capture down to individual point defects[4], as a mechanism for electrical readout of spin qubits[5], or as route to expose non-fluorescent charge emitters in a solid[6]; spin-to-charge conversion has also emerged as a strategy to enhance spin readout sensitivity[7-9], particularly important in applications to metrology[10,11]. Since thermal activation of deep donors or acceptors is negligible in insulators, excess trapped charges have long lifetimes, especially at low and moderate concentrations where tunneling processes between neighboring defects are strongly suppressed. Color centers can therefore be seen as atomic-size optical memories, a notion already exploited to demonstrate reversible, three-dimensional data storage[12-14].

An important hurdle impacting color center charge control is light diffraction, limiting spatial discrimination to hundreds of nanometers. Color center imaging with precision down to a few nanometers can be attained via super-resolution techniques such as stimulated emission depletion[15] (STED) or stochastic optical reconstruction microscopy[16,17] (STORM). Unfortunately, applying these methods to sub-diffraction charge control is impractical, either because the illumination intensity is too weak to induce controllable charge state conversion — the case for STORM — or because the finite, diffraction-limited size of the STED beam unavoidably alters the charge state of color centers proximal to a target[18-20]. Several STED-based approaches to sub-diffraction optical lithography have been demonstrated in the recent past[21-26] but the underlying mechanism — generically based on the ability to induce a photo-chemical reaction with high spatial selectivity — is non-reversible, requires careful alignment of an ancillary, doughnut-shaped beam, and is often limited to two dimensions.

Here, we implement charge control protocols in NV centers in diamond under cryogenic conditions, where narrow optical transitions reveal the fine structure in the ground and first excited states[27]. We first focus on small sets of color centers within the same diffraction-limited volume and show charge state preparation of individual NVs using selective, local-field-shifted optical transitions. Switching to denser ensembles where individual control becomes impractical, we nonetheless demonstrate charge state writing and readout of frequency-binned NV subsets within the same focal volume, which we then exploit to show multiplexed data storage on the optical plane.

**NV-selective imaging**

Figure 1a shows the energy level structure of the negatively charged NV, featuring a triplet $A_2$ ground state with spin sublevels $m_S = 0, \pm 1$, and an excited state manifold combining a spin triplet and an orbital doublet $E_{x,y}$[27,28]. We obtain excitation spectra by recording the Stokes-shifted NV⁻ emission as we sweep the frequency of a 637-nm, narrow-band laser. The level structure of Fig. 1a leads to six optical resonances connecting ground and excited state levels with the same spin projection. In practice, however, state hybridization in the excited manifold leads to spin flips that render the NV non-resonant, hence preventing the observation of most transitions in single-wavelength excitation experiments[27].

One route to circumvent this problem is to accompany laser illumination with microwave (MW) excitation resonant with the two NV spin transitions of the ground state triplet[11,27]. Here we follow a different route where pulses of resonant optical excitation near 637 nm alternate with intervals of 532 nm laser light so as to periodically re-pump NV⁻ into the $m_S = 0$ state and recharge NV⁰ into NV⁻ in the event of ionization. We demonstrate this strategy in Fig. 1b where we investigate a set of NVs in a type 2a [100] diamond (below referred to as Crystal A). Standard confocal microscopy under green illumination (see central image) reveals a discrete collection of NV centers, whose optical spectra we record using the protocol in the upper diagram. Importantly, all NVs exhibit strain- or electric-field-shifted optical resonances[29,30] that we can address individually to reconstruct NV-selective images. We illustrate the idea for four different emitters in the set, in each case configuring the detection protocol so that the 637-nm laser is resonant with an individual $S_z$ ($m_S = 0$) transition. Similar to experiments with SiV centers[31], the images we obtain highlight almost exclusively the chosen NV, a combined consequence of the laser selectivity and the broad spectral heterogeneity produced by local electric and crystal strain fields (Supplementary Material, Section 1). Remarkably, resonant optical excitation occasionally reveals NVs not observable via green illumination (see, e.g., upper right insert in Fig. 1b), partially a consequence of the NV fluorescence sensitivity to the incoming beam polarization (Supplementary Material, Section 2).

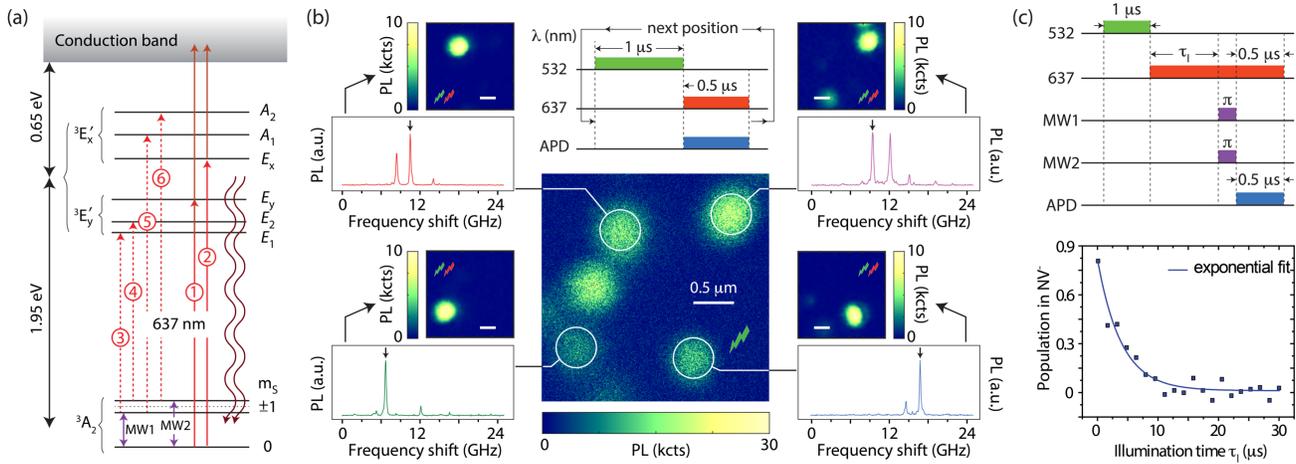

**Figure 1 | Optical spectroscopy and charge control of NV⁻ centers under cryogenic conditions.** (a) Energy level diagram of NV⁻. Light red arrows (solid and dashed) indicate optical transitions around 637 nm between levels in the ground and first excited manifolds; dark red arrows indicate ionization photons, and wavy arrows denote emitted photons. (b) (Main) Scanning confocal image under green excitation of a section of the crystal featuring multiple NVs. (Side inserts) Optical spectra of circled NVs in the set upon application of the protocol in the upper diagram using red illumination of variable frequency; here (and everywhere else unless noted), the horizontal axis is a frequency shift relative to 470.470 THz. For each case, we obtain an NV-selective image using the same protocol but with the 637-nm laser tuned to one of the $S_z$ transitions (indicated by an arrow in each spectrum); only the resonant NV⁻ is visible in the images. The laser powers are 1.6 mW and 2 μW at 532 and 637 nm, respectively. (c) (Top) NV⁻ ionization protocol under strong optical excitation (210 μW) at 637 nm. MW1 (MW2) denotes MW excitation resonant with the $m_S = 0 \leftrightarrow m_S = -1$ ($m_S = 0 \leftrightarrow m_S = +1$) transition in the ground state triplet; the duration of the π-pulses is 100 ns. (Bottom) Relative NV⁻ charge state population as a function of the ionization interval $\tau_I$ for a representative NV in the set. All experiments are carried out at 7 K. PL: Photoluminescence. a.u.: Arbitrary units.

To gain single NV charge state control, we bring the laser frequency on-resonance with one of the cycling transitions ($m_S = 0$), and increase the power so that NV⁻ ionization takes place on a time scale faster than the inverse spin flipping rate (upper schematic in Fig. 1c). We confirm this condition through the application of inversion MW pulses resonant with each of the two NV⁻ spin transitions at the applied magnetic field ($B = 4.51 \pm 0.05$ mT, see Methods); negligible fluorescence upon spin inversion signals NV⁻ ionization.

**Sub-diffraction ionization of individual NVs**

We leverage our charge-state-control approach in Fig. 2a where we zero-in on a diffraction-limited site in the crystal hosting a pair of NVs, a composition we infer from the multi-peak structure of the optically-detected magnetic resonance (ODMR) spectrum (upper right-hand insert in Fig. 2a). We separately visualize each NV in the pair via a spatial reconstruction obtained from repeatedly imaging the fluorescence pattern from either emitter under selective excitation, and subsequently fitting the result with a Gaussian point spread function[32] (left-hand side insert in Fig. 2a). When combined, the above data set reveals the presence of two NVs — here referred to as $NV_A$ and $NV_B$ — oriented along non-equivalent crystalline axes and separated by an in-plane distance of 90 nm, below optical resolution (~300 nm in the present case).

Optical spectroscopy of the same crystal site shows a collection of resonances, each associated with one emitter or the other (lower right-hand side plot in Fig. 2a). To separate these resonances, we implement a detection protocol identical to that in Fig. 1b, except that we intercalate MW spin inversion pulses between the green and red laser excitation intervals selectively acting only on one of the two emitters (upper two traces in the optical spectra of Fig. 2a).

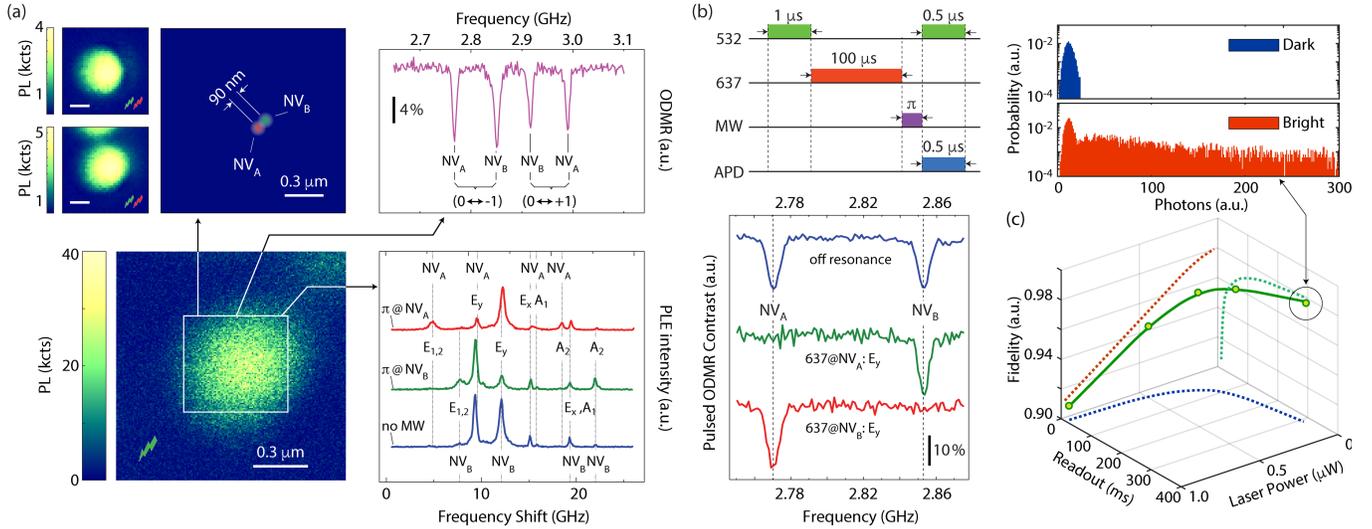

**Figure 2 | Sub-diffraction NV⁻ ionization and readout.** (a) (Lower left) Confocal microscopy under 532 nm illumination of a diamond site containing two NVs. (Upper left inserts) NV selective imaging using the protocol in Fig. 1b and reconstructed sub-diffraction map as derived from Gaussian point-spread-function fits; colored circles indicate the mean positions and standard deviations obtained from comparing multiple images over time. (Upper right insert) ODMR spectrum under 532 nm excitation; the four resonances we observe confirm the presence of two differently oriented NVs. (Lower right insert) NV⁻ optical spectroscopy in the absence of MW (blue trace) or upon intercalating a MW π-pulse between the laser pulses in the protocol of Fig. 1b; the MW frequency is chosen resonant with the spin transitions of $NV_A$ or $NV_B$ (green and red traces, respectively). (b) (Top) NV-selective ionization protocol; for inspection, we reconstruct an ODMR spectrum by varying the MW frequency of the π-pulse (set to probe a band around the $|0\rangle \leftrightarrow |-1\rangle$ transition). (Bottom) We ionize $NV_A$ or $NV_B$ by bringing the 210-μW red laser on-resonance with the $E_y$ transition of either color center (green and red traces, respectively); we recover the full ODMR spectrum when the 637-nm laser is tuned away from all transition frequencies. In (a) and (b), spectra have been displaced vertically for clarity. (c) Measured charge state readout fidelity for $NV_A$ a function of the resonant laser power and readout time. The solid line is a guide to the eye and the dashed lines represent projections onto each plane. The top insert shows the measured photon count probability distributions upon probabilistic charge state initialization into NV⁻ and NV⁰ (bright and dark states, respectively); the extracted readout fidelity in this case is 98% (Supplementary Material, Section 3).

Comparing the system response in the absence of MW excitation (lower trace) allows us to assign all optical resonances in the pair.

To demonstrate sub-diffraction NV⁻ charge control, we make the red laser frequency resonant with one of the $m_S = 0$ transitions in either NV and adjust the illumination intensity and duration to quickly produce ionization (pulse diagram in Fig. 2b). We probe the ensuing charge state of the pair via an ODMR protocol in which we measure the system fluorescence as we vary the frequency of a MW spin inversion pulse. As shown in the main plot of Fig. 2b, we identify complementary ODMR spectra where the observable dip is contingent on the chosen narrow-band excitation frequency of the 637-nm laser (mid and lower traces); comparison with the ODMR spectrum under non-resonant red illumination — where both dips are visible, upper trace — flags NV-selective charge control. Note that since each ODMR trace necessarily involves multiple repeats, the plot in Fig. 2b simultaneously demonstrates our ability to initialize, write, and readout the charge state of each NV a virtually unlimited number of times.

While resonant excitation NV⁻ ionization deterministic (Fig. 1c), the uncertainty associated with each individual charge state readout is a priori unclear. To address this question, we implement a 'single-shot' charge state detection protocol and determine the readout fidelity from the photon count distributions we measure upon NV charge state preparation. The histograms in Fig. 2c show a representative example; note that probabilistic initialization into NV⁻ (~80% in this case) leads to overlap, which, nonetheless, we can deconvolve with the aid of post-selection (see Methods and Supplementary Material, Section 3). In the regime of low laser powers, we attain fidelities reaching up to 98%, limited by spectral diffusion[33]. Increasing the illumination intensity to shorten the readout time also has an impact on the resulting readout fidelity, thus leading to a complex interplay. We capture this multi-parameter response through a three-dimensional plot that takes into account the tradeoff between laser power and readout duration. Remarkably, we find that the readout fidelity remains above 90% even for the shortest readout times (2 ms). Importantly, the observed response is largely a function of the photon collection efficiency — reduced by the long working distance of our objective, see Methods — implying there is still room for improvement.

We integrate charge state writing and single-shot readout in Fig. 3 where we investigate the response of an NV cluster confined within the focal volume (approximately 0.3×0.3×1 μm). Combining frequency selective NV imaging and optical spectroscopy under resonant microwave excitation, we identify four distinct NVs whose optical transitions we can map to those expected in the presence of local strain and electric fields (Figs. 3a through 3d, see also Supplementary Material, Section 4). We demonstrate selective charge control and non-destructive readout of the NV cluster in Fig. 3e: Starting from a state where only $NV_E$ is neutral (trace ①), we successively bring the red laser frequency on resonance with $NV_D$ and $NV_C$ transitions to induce selective ionization, two charge state configurations we confirm by reconstructing in each case the cluster's optical spectrum (respectively traces ② and ③). Note that unlike the case in Fig. 2, the entire process takes place in the absence of green illumination — and hence without resorting to a charge state reset; this non-destructive read-out of the cluster alleviates the detrimental effect of spectral diffusion and provides a near-systematic pathway to discriminating between contributions from individual NVs in crowded spectra. Additional observations at half-hour intervals confirm the final charge state configuration in the set, though multiple readouts ultimately lead to ionization as shown in trace ⑧ (see also Supplementary Material, Section 4).

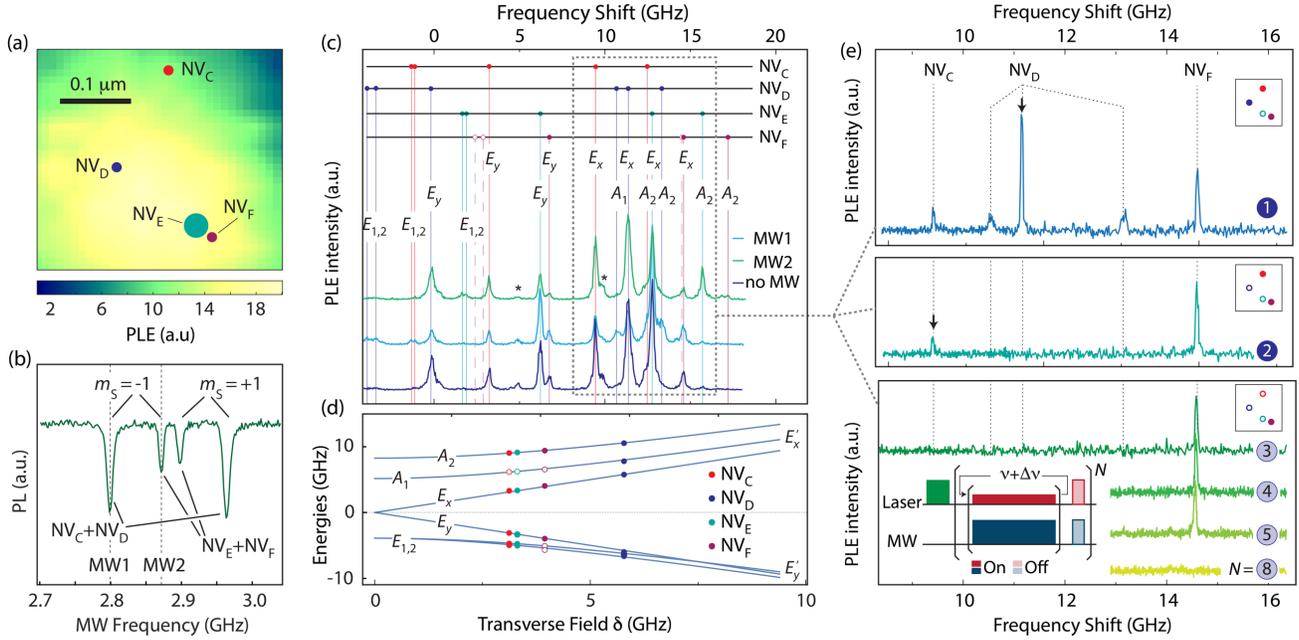

**Figure 3 | Charge manipulation and non-destructive optical spectroscopy of a four-NV sub-diffraction cluster.** (a) Confocal image of an NV cluster sharing the same diffraction-limited volume. The superimposed circles show the mean positions of each NV as determined from selective excitation; diameters indicate the uncertainties. (b, c) ODMR and optical spectroscopy of the four-NV cluster using the protocol in Fig. 1b; blue, light-blue, and green traces illustrate the cluster response in the absence of MW or upon intercalating MW inversion pulses resonant with different ODMR transitions; asterisks mark un-assigned PLE peaks, perhaps associated to a fifth NV in the cluster. Empty circles denote weak or unobserved transitions. (d) Calculated energies in the NV excited manifold as a function of the transverse electric/strain field. Full circles indicate the measured energies for each NV in the cluster as derived from (c). (e) Non-destructive optical spectroscopy of the cluster (see schematics in the lower left where ν denotes laser frequency and Δν is the step during the spectral reconstruction). In this illustration, green illumination (1 μW, 10 s) initializes $NV_C$, $NV_D$, and $NV_F$ into the negatively charged state while $NV_E$ remains neutral and is thus unobservable (trace ①). Traces ② and ③ show the resulting spectra upon ionizing first $NV_D$, then $NV_C$ with the laser tuned to the transition indicated by the arrow; full (empty) circles in the upper right inserts indicate negative (neutral) NVs in the cluster using the color code in (a). Trace ④ and ⑤ are successive non-destructive readouts of the same charge configuration; trace ⑧ shows the result of unintended ionization of $NV_F$ during readout. Throughout these spectra, the red laser power is 5 nW during readout and 50 μW during the 200-ms-long ionization pulses; reconstructing one spectrum typically takes 3 minutes. In (c) and (d), the laser frequency shifts are relative to 470.480 THz.

Along with the opportunities inherent to individual NV charge state control, the experiments in Fig. 3e also expose some of the present limitations. For example, we find that spectral diffusion — already present at the level of individual NVs — is a bit greater in the cluster (Supplementary Material, Section 1). Additionally, charge initialization into $NV^-$ — here relying on probabilistic recombination of $NV^0$ under green illumination[34] — is less likely in the cluster, and shows characteristic rates that change from one NV to the other. Further work will be needed to understand and ultimately mitigate this problem, perhaps by exploiting the narrow $NV^0$ PLE lines[35] to selectively induce one-way recombination into $NV^-$ under 575-nm illumination.

## Multiplexed storage in NV ensembles

Besides featuring a more complex charge dynamics, denser NV sets necessarily lead to congested optical spectra where individually addressing all NVs ultimately becomes impractical. To explore the response of larger NV ensembles, we now turn to Crystal B, a chemical-vapor-deposition-grown diamond hosting an NV concentration of ~30 ppb, orders of magnitude greater than in Crystal A. Given the sub-diffraction average distance between NVs — of order 60 nm at this concentration — standard confocal microscopy cannot discriminate individual emitters, instead yielding featureless images that only capture the mesoscale, naturally-occurring variations in NV number. On the other hand, strain and electric fields present in this diamond lead to a broad distribution of NV$^-$ optical resonances, which potentially allow one to generalize the experiments of Figs. 2 and 3 to the ensemble. Indeed, photoluminescence excitation spectroscopy — obtained from a spatial average over a (50 μm)$^2$ area in an inner diamond plane — reveals for this crystal a 100-GHz-broad, Gaussian distribution of NV$^-$ optical resonances (Fig. 4a). A natural question is, therefore, whether one can multiplex the information content of an optically defined pixel by selectively ionizing multiple subsets of NVs within a diffraction-limited focal volume. We note that the answer is not immediate, as it is not a priori clear that the impact of narrow-band illumination — ionizing the subset of NVs with optical transitions at the laser frequency — can be confined to a spectral range smaller than the full heterogeneous linewidth (the converse would make crosstalk unavoidable). Further, direct examination is needed to assess the effect of carrier diffusion and recapture, potentially altering — during writing and/or readout — the charge state of previously encoded NVs (Supplementary Material, Section 5).

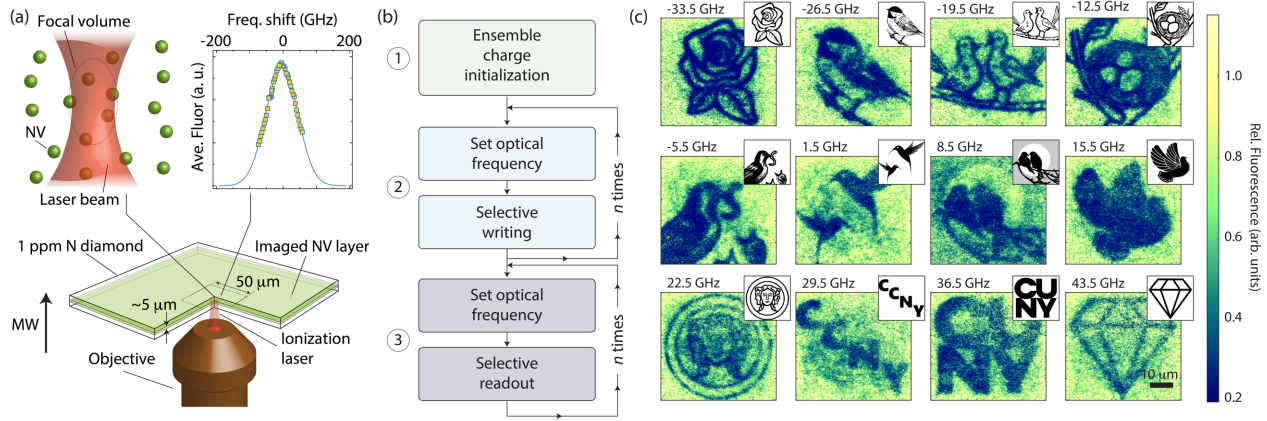

**Figure 4 | Multiplexed data storage.** (a) We study a [100] type 1b diamond with NV content greater than in crystal A. Confocal imaging and optical spectroscopy in a section of an inner, 5-μm-thick plane reveal an average inter-NV separation below diffraction as well as a broad, strain-induced distribution of optical transitions (upper right insert). (b) Schematics of the data storage protocol. We use a 532-nm scan (61 μW) to initialize NVs into the negatively charged state (step ①) followed by band-selective 637-nm illumination (20 μW) to write $n$ charge-encoded, spatially overlapping patterns (step ②); we implement a similar protocol for data readout (step ③) except that we reduce the laser illumination power to 2 μW so as to avoid NV$^-$ ionization. (c) Frequency-selective readout of the in-plane, charge-encoded data set; the pixel size is 0.36 μm$^2$, and the readout time per pixel is 1 ms. The order in the image reconstruction is left-to-right, top-to-bottom, hence leading to a gradual loss of contrast (Supplementary Material, Section 5). In (a) and (c) frequency shifts are measured relative to 470.48039 THz; no external magnetic field is present and MW excitation at the NV zero-field splitting frequency (2.877 GHz) accompanies resonant optical excitation during writing and readout.

That these problems can be circumvented is shown in Fig. 4c, where we address groups of NVs with optical transitions centered around different optical frequencies spaced every 7 GHz; to more easily assess the technique's fidelity, we structure our storage protocol so as to create a collection of spatial patterns that we can visually examine upon image reconstruction. The result is a collection of 12 different charge-encoded patterns simultaneously coexisting within the same crystal plane with minimal mutual interference. Given the comparatively narrow linewidth of an individual optical transition (~100 MHz), one can conceivably increase the level of multiplexing by reducing the bandwidth of all frequency windows. Unintended crosstalk, however, becomes apparent for frequency steps narrower than 7 GHz, hence imposing a practical limit (Supplementary Material, Section 5).

Additional constraints derive from the finite NV content, particularly if the spatial distribution of color centers on the plane is to remain reasonably uniform. While greater NV concentrations can potentially mitigate the problem, a critical threshold is to be expected as metastable charge states gradually succumb to Fermi statistics. From prior observations[12], we tentatively set the upper bound for NV charge metastability below 100 ppb, though we caution this value can vary largely depending on the concentration of co-existing charge traps, most notably vacancies as well as nitrogen and silicon impurities. For example, while charge states are long-lived in Crystal B at cryogenic temperatures, we observe some data loss after a four-day thermal cycle from 4 K to room temperature and back (Supplementary Material, Section 6). Interestingly, the NV charge state has been seen to remain stable in diamonds with similar NV content (but different composition) over periods exceeding a week at room temperature[12], hence indicating cryogenic temperatures may ultimately not be required for data storage.

**Conclusions**

In summary, the narrow zero-phonon lines of NVs at cryogenic temperatures combined with naturally occurring electric and strain fields in the host diamond crystal allow us to attain charge control of individual color centers with sub-diffraction resolution. Generalizing to ensembles of NVs, we demonstrate data multiplexing within the same optical plane, which, as shown previously[12], can be extended to three dimensions, hence boosting the storage capacity of a crystal multiple fold.

Extensive additional research is mandatory before these findings can translate into a viable data storage technology, although several features make the present approach of potential interest. For example, there is arguably no limit on the number of photo-activated cycles of ionization and recombination an NV can undergo; further, since charge control relies on resonant optical excitation the total amount of laser energy required per writing operation is low (of order 2 nJ for a beam intensity of 0.2 MW cm$^{-2}$), especially if compared to STED-based lithography methods (where the typical inhibition beam energy is of order 10 MW cm$^{-2}$ for a writing energy of ~1 µJ at best[21], typically more).

Although rewriting a diffraction limited volume as presently demonstrated requires re-initialization into NV$^-$, selective recombination from a state where all NVs are neutral could potentially be attained via resonant, narrow-band excitation of the strain-induced transitions around 575 nm[36] (unfortunately unavailable in our present experimental setup). Extending our proof-of-principle demonstrations, future work may combine resonant optical excitation with super-resolution methods to better understand the interplay between strain fields, spatial proximity, and selective optical addressability. Along related lines, material growth techniques could be adapted to produce crystals where the composition and strain fields are chosen to optimize NV charge state stability and spectral dispersion[37]. Attaining individual color center control in these ensembles could prove rewarding, especially when considering the potentially large storage capacity of a small crystal, even at low color center concentrations. For illustration, an NV content of ~1 ppb — comparable to that attained locally in Fig. 3 — would correspond to a capacity of 2 petabits/inch$^3$, which compares favorably with alternative optical storage technologies (Supplementary Material, Section 8).

Given the ubiquity of color centers featuring alternative charge states, this method could find extensions through the use of emitters other than the NV (e.g., the silicon-vacancy center in diamond) or in other material platforms (such as silicon carbide, hexagonal boron nitride, or rare-earth-hosting garnets). Adding to wide-bandgap materials, an intriguing crystal host is silicon, a system where readout could be facilitated through the use of integrated photonic structures.

Besides applications to data storage, the ability to selectively address optical transitions of color centers separated by sub-diffraction distances could find use in the investigation of carrier transport between physically distant color centers[38,39] or to study correlated electric noise[40]. Similarly, recent observations in diamond-hosted NV and SiV centers (respectively, Refs. 4 and 41) suggest the use of resonant near-ionizing excitation to induce bound-exciton Rydberg states; the delocalized, large-radius nature of these orbitals could be exploited as a platform to induce dipole blockading between proximal color centers, a step towards replicating atomic physics phenomena in the solid state[42].

## Methods

### Experimental setup

Throughout our experiments, we use a home-built confocal microscope[43] with excitation in the green (532 nm) and the red (637 nm). We merge the laser beams into a single excitation path with a 605-nm short-pass filter onto a single mode fiber. We produce laser pulses with the aid of 10-ns-rise-time acousto-optic modulators, and detect NV fluorescence via a single-photon avalanche-photo-detector. Long-pass filters and a 650-nm dichroic mirror allow us to reject laser excitation leaking into the detection path. We use a permanent magnet to create a small magnetic field (4 mT) close to the vacuum chamber, and a cryo-workstation (Montana Instruments) to control the sample temperature down to 5 K. Inside the vacuum chamber sits an air objective (100× Zeiss Epiplan Neofluar with a numerical aperture of 0.75). We use a tunable laser diode (Toptica DL pro HP 637) for resonant excitation, which we stabilize (±200MHz absolute accuracy) with a wavemeter (High-Finesse WS/6-200). We control the NV electron's spin via pulsed microwave excitation produced by two signal generators (Rhode&Schwarz SMB100A and Stanford Research Systems SG386) and gate either signal via RF switches from Minicircuits; a Pulseblaster TTL-pulse generator controls the timing of all pulses in our protocols. A 25 µm copper wire overlaid on the sample serves as the MW source in protocols demanding spin manipulation.

### Sample details

Experiments on single NV centers use a (100), electronic-grade diamond crystal from Element6. Experiments on ensembles, on the other hand, rely on a (100) diamond from DDK with approximately 1 ppm of nitrogen and estimated NV content of 30 ppb. Unlike other similar crystals from the same vendor, this particular crystal shows no traces of SiV centers.

### Measurement protocols

*Individual NV ionization:* For NV-selective ionization measurements, we charge-initialize NVs via a 1.6-mW, 1-µs-long laser pulse (532 nm) followed by a 210-µW, 100-µs light pulse (637 nm) for selective ionization. Both NVs exhibit similar Rabi frequencies of 9.09 MHz for the chosen MW field strength at their respective $|0\rangle \leftrightarrow |-1\rangle$ transitions in the ground triplet. To reconstruct the ODMR spectra in Fig. 2b, we monitor the system photoluminescence as we sweep the MW frequency for a variable (but fixed) ionization pulse wavelength. The increase in spin contrast for the non-ionized NV stems from a decrease in fluorescence upon ionization of the other NV. Throughout these experiments, the magnetic field has a magnitude $B = 4.51 \pm 0.05$ mT; from the observed $|0\rangle \leftrightarrow |\pm 1\rangle$ ODMR resonance dips, we conclude[44] $B$ forms angles $\theta_{NV_1} = 28.8 \pm 1.14$ deg. and $\theta_{NV_2} =$

105.9 ± 0.19 deg. with respect to NV$_1$ and NV$_2$, respectively.

*Charge state readout fidelity of individual NVs:* To characterize the charge state readout fidelity, we use an initialization-readout sequence comprising a green laser pulse (100 µW for 10-1000 µs), a short red laser pulse for data post-selection (100-1000 µs) and a longer red laser pulse for charge read-out. The same sequence is implemented only with red laser to determine the NV$^0$ state readout fidelity. For a given red laser power, we choose the post-selection pulse length such that ionization can be neglected compared to other readout errors. The post-selection criteria are then chosen to bring the initialization error to the same level, keeping only 8-25% of the cases. Finally, the photon threshold between bright and dark read-out is taken such as to simultaneously maximize fidelity of the bright state readout (post-selected data) and dark state readout (red-only sequence). For a given red laser power (50-500 nW) we choose the duration of the red laser readout pulse (20-400 ms) such as to maximize those fidelities, with durations generally close to the defect average ionization time. These parameters are then used on the non-selected data to determine the bright state initialization fidelity (79% in the single NV case). In order to build enough statistics, sequences are typically repeated 10000 and 5000 times for the bright and dark state respectively.

*NV ensemble imprint:* To initialize the background to a bright NV$^-$ state, we perform a 61-µW, 532-nm scan with 1 ms dwell time per pixel. Then, we use a 20-µW, 637-nm laser to imprint patterns with dwell times varying from 2 ms to 500 ns for grayscale images. For multiplexed imaging, NV$^-$ bright state initialization is performed once. Following each imprint, we use the wavemeter to stabilize the laser frequency at the next desired frequency. After imprinting all desired images, we use a weak 2-µW laser for readout at each chosen frequency. All images contain 300×300 pixels in a 50×50 µm$^2$ window. Throughout these experiments we apply no external magnetic field. To prevent spin depletion, we accompany resonant optical excitation at 637 nm with a strong MW excitation at 2.877 GHz that equilibrates spin populations between all levels in the ground triplet. Continuous microwave excitation raises the sample temperature by < 1 K.

*Thermal cycling:* We imprint and readout images following the protocol described in the main text and sub-section above. Due to mechanical drift of the system during the temperature cycling process (in the order of tens of microns), we use the X-shaped image (at a frequency of 470.4495 THz) as a "marker" for localization of the region of interest. We perform several weak, 300-nW scans at the "marker" frequency to re-position the sample back into the original focal plane and window of interest. The diamond crystal remains in the dark throughout the thermal cycling process. Additional details can be found in Section 6 of the Supplementary Material.

## Data availability

The data that support the findings of this study are available from the corresponding author upon reasonable request.

## Acknowledgments

The authors acknowledge useful discussions with D. Irber, F. Reinhard, and A. Lozovoi. R.M. and C.A.M acknowledge support from the National Science Foundation through grant NSF-1914945; T.D. and C.A.M. acknowledge support from NSF through grant NSF-2216838. R.M. acknowledges support from NSF-2316693. All authors acknowledge access to the facilities and research infrastructure of the NSF CREST IDEALS, grant number NSF-2112550.

## Author contributions

R.M., T.D. and C.A.M. conceived the experiments. R.M. and T.D. conducted the experiments and analyzed the data with C.A.M's assistance. C.A.M. supervised the project and wrote the manuscript with input from all authors.

## Competing interests

The authors declare no competing interests.

## Correspondence

Correspondence and requests for materials should be addressed to C.A.M.

# Table of Content

## 1. Low temperature optical spectroscopy of individual NVs

For our experiments, we use a custom-made confocal microscope[1] based on green laser excitation (532 nm). The system has been adapted to also accommodate a 500-kHz linewidth laser (Toptica) tunable over a few nanometers around 637 nm; we lock the laser frequency with the aid of a wavemeter (High Finesse). Linear polarizers along with broadband half- and quarter-wave plates preceding the objective control the input polarization of both laser beams. We manipulate the NV spin through microwave (MW) pulses stemming from a thin copper wire (25 µm diameter) overlaid on the diamond surface. We use a cryo-workstation (Montana Instruments) to attain low-temperatures down to 4 K; the exact crystal temperature — typically in the 5 to 7.5 K range — varies depending on the operating conditions, most notably the laser intensity and microwave load. Optical excitation and fluorescence collection rely on a 0.75 numerical aperture objective with a working distance of 4 mm; the objective sits inside the vacuum chamber of the cryostat but does not have thermal contact with the main cold plate, hence reducing the amplitude of the thermal cycle it experiences during cool-downs (and correspondingly extending its lifetime). We collect the photon emission into a multi-mode fiber, which also serves as the confocal pinhole; we use a pair of single-photon avalanche photodetectors (Excelitas Tech.) in the Hanbury Brown-Twiss geometry to detect photoluminescence from individual emitters.

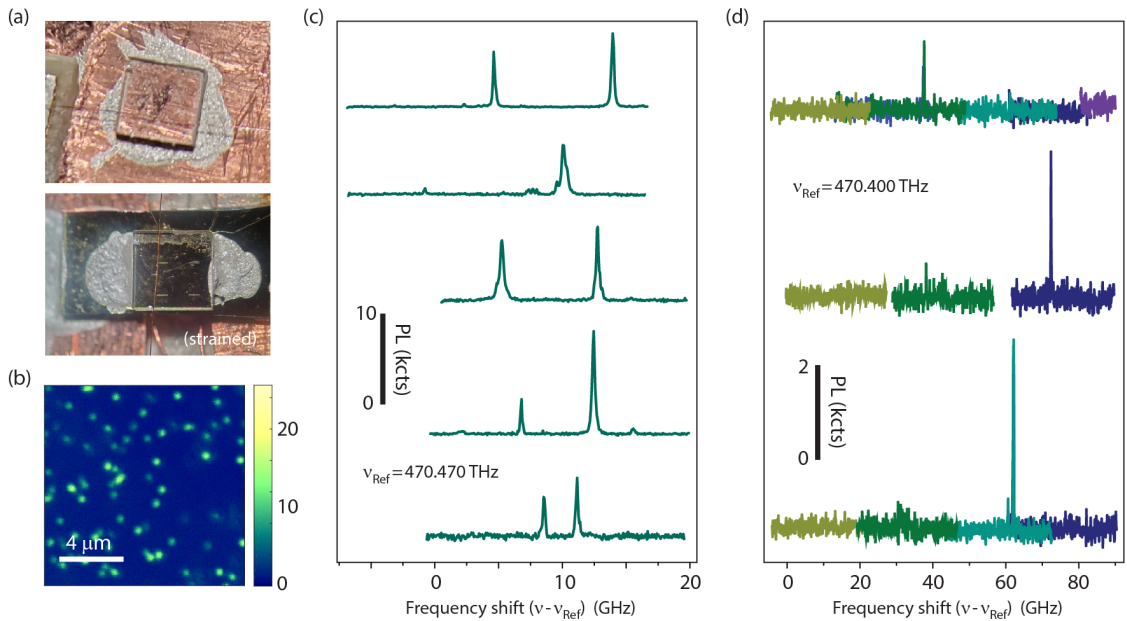

**Fig. S1: Cool-down-induced strain.** (a) We use silver paste to establish good thermal contact between the diamond crystal and its substrate while holding the sample in place. Differences between the thermal expansion coefficients of diamond and all other materials in the mount lead to strong crystal strain upon cool-down if the mount is too rigid (lower image). This problem can be mitigated by letting a minimal amount of silver paste creep into the interface between the diamond and the substrate (upper image). (b) Representative confocal image of the diamond in (a) (crystal A) under green illumination. (c) Example PLE spectra from randomly selected NVs in the "unstrained" (i.e., minimally strained) crystal. (d) Same as in (c) but for the case where the crystal is rigidly mounted ("strained" crystal in (a)); colors indicate different frequency scans of the same NV after manually retuning the red laser (the hop-free range is ~25 GHz). Compared to (c), we find that most NVs (~75%) do not show PLE resonances in the 100 GHz range we probed (not shown in the graph); those that do, exhibit lower photoluminescence (PL) and their resonances are more sparsely distributed. In (c) and (d), the crystal temperature is 7.5 K.

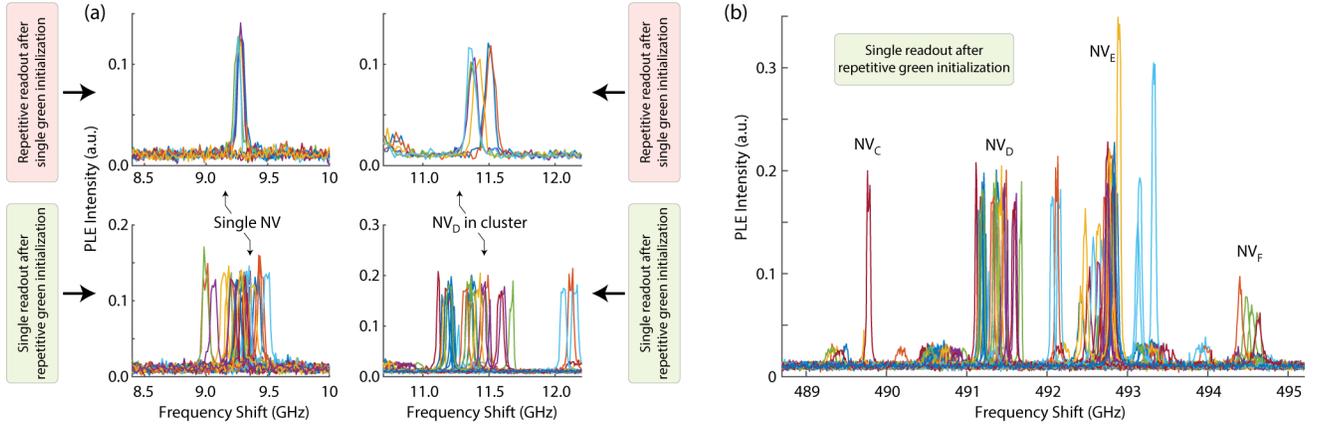

**Fig. S2: Spectral diffusion of PLE resonances.** (a) Representative PLE spectroscopy upon multiple readout of one of the optical resonances for an individual NV and for $NV_D$ in the cluster of Fig. 3 in the main text (left and right columns, respectively). The upper row corresponds to spectra obtained via red excitation after a one-time charge initialization with green light; the spectra in the lower row also derive from red readout but green light is used to charge initialize the NV prior to each pass. (b) Same as in the lower left plot in (a) but extended throughout the $E_x$ manifold in the cluster.

To ensure good thermal contact, the diamond sits on a copper or sapphire substrate, in turn directly connected to the cryostat's cold plate through gold thermal braids. Since the objective is in a horizontal position[1] (i.e., the beam propagates parallel to the optical table as it hits the sample surface), we make use of silver paint (Silver Conductive Adhesive 503 from Electron Microscope Science) to hold the diamond in place (Fig. S1a). The paint amount, location, and viscosity have a large impact on the accumulated crystal strain upon cool-down, and hence on the optical response of the NVs. For example, Figs. S1b through S1d show representative confocal images and photoluminescence excitation (PLE) spectra from Crystal A, an electronic grade diamond from E6. In a cool-down captured by Fig. S1c, about 90% of the NVs we inspected show multiple PLE resonances within the same spectral range, and the photoluminescence is comparatively high; by contrast, slight changes in the sample mount makes most NVs dark (roughly 75% from a total of 20), and those that fluoresce tend to display dim spectra with resonances scattered over a much wider range. We interpret these observations as the result of cool-down-induced crystal strain, which, in turn, we relate to the silver bond with the substrate. Specifically, we find that a geometry where a small amount of silver paint creeps between the diamond back-surface and the substrate brings crystal strain to a minimum; fresh (i.e., solvent-rich) paints work best (Fig. S1a).

Similar to prior observations[2], we find that individual NVs in Crystal A undergo significant spectral diffusion, most likely associated with a redistribution of trapped carriers upon photo-activation, diffusion, and re-capture[3-5] (Fig. S2). Indeed, the use of green excitation preceding red readout introduces comparatively greater fluctuations (compare upper and lower plots in Fig. S2a), which can be attributed to more effective ionization and/or recombination processes at shorter wavelengths. Interestingly, spectral fluctuations are comparatively greater for NVs in the cluster, leading to short-term frequency drifts of order 500 MHz (compare left and right plots in Fig. S2a, see also Fig. S2b). The spectral range stemming from these fluctuations — likely indicative of defect aggregation at this site — is about 4 times greater than reported in early studies[2], but considerably smaller than that produced by mechanical strain. This said, we cannot rule out electric fields from charges in deeper traps — insensitive to green or red excitation — as ultimately responsible for the observed spectral inhomogeneity. Unlike crystal strain, fields stemming from neighboring traps should be comparatively more local, implying that a statistical study of PLE spectra from proximal NVs — e.g., separated by less than 100 nm — should allow one to gauge the relative impact of either source on the observed spectral heterogeneity. We present initial results along these lines in Section 5 but postpone an in-depth investigation to future work.

## 2. NV-selective optical microscopy

Our microscope uses linear polarizers as well as half- and quarter-wave plates to independently control the polarization of both incoming beams; to reconstruct a fluorescence image, we use the mirrors of a galvo system to scan the focal plane. As shown in the schematics of Fig. 1b in the main text, we implement NV-selective images with the aid of a protocol where fluorescence readout takes place selectively under red light in a train that alternates 532- and 637-nm pulses. Assuming red laser tuning to one of the $S_z$ transitions (see Fig. 1a in the main text), green illumination re-pumps NV$^-$ back into the $|m_S = 0\rangle$ spin state in the ground state triplet, and thus compensates for inevitable spin depletion during resonant excitation[6]; regular green illumination also prevents trapping of the NV in the neutral charge state[7] (dark under 637-nm light) upon unintended red ionization. Unlike in PL images recorded under green illumination — presumably revealing all NVs in the scanned focal plane — only few emerge under the green/red sequence, namely those featuring optical transitions at or near the chosen red laser frequency.

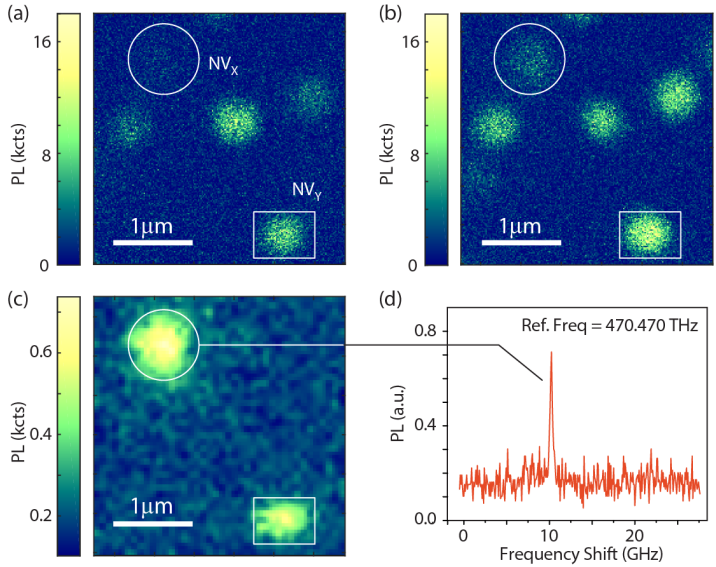

**Fig. S3: Green versus resonant NV imaging.** (a, b) Confocal images of the same area in crystal A; in both cases we use green illumination, but the beam's linear polarization is differently oriented. We see a change in the relative brightness of all NVs including $NV_X$, virtually invisible in (a); its brightness, however, remains faint, even upon optimization of the green beam polarization as shown in (b) (circle in the upper left corner). (c, d) Resonant confocal imaging of the same area using the protocol in Fig. 1b of the main text under similar conditions. The laser frequency is 470.47985 THz, which is resonant with the $S_z(E_y)$ transition of $NV_Y$ (white square in the lower right corner). The laser power is sufficiently strong to simultaneously excite $NV_X$, featuring a nearby optical resonance (spectrum on the right-hand side). Interestingly, $NV_X$ shows up prominently in the resonant image, even though its presence is barely noticeable under green excitation.

Intriguingly, we find in rare occasions that green excitation fails to excite NVs efficiently. Fig. S3 provides a first illustration: Comparison of the fluorescence images obtained under green or red illumination (respectively S3a and S3c) shows that $NV_X$ — a color center virtually invisible during the green scan, circle in the upper right corner of the image — features, nonetheless an unmistakable fluorescence signal under red readout. Further, while the NV excitation efficiency is known to depend on the relative orientation of the plane of polarization, the fluorescence from $NV_X$ remains faint even for optimal beam polarization. We observe similar phenomenology in the upper right (lower left) insert image of Fig. 1b (Fig. 2a) in the main text. We presently ignore the mechanisms underlying the observed NV response though we hypothesize differences in the response may partly stem from a disparity and/or slight misalignment in the focal volumes associated to the red and green beams. Additional work will be needed, however, to gain a fuller understanding.

### 3. Charge readout fidelity of individual NVs

While a sufficiently long exposure to red laser light guarantees $NV^-$ ionization (or, equivalently, initialization into $NV^0$), green illumination (which excites non-resonantly both $NV^0$ and $NV^-$) leads to $NV^-$ formation probabilistically[8] (the yield is typically 75% to 80% depending on temperature). This limitation creates complications when quantifying the charge readout fidelity in individual NVs, as one must separate events of green excitation that result in the unintended formation of NV$^0$.

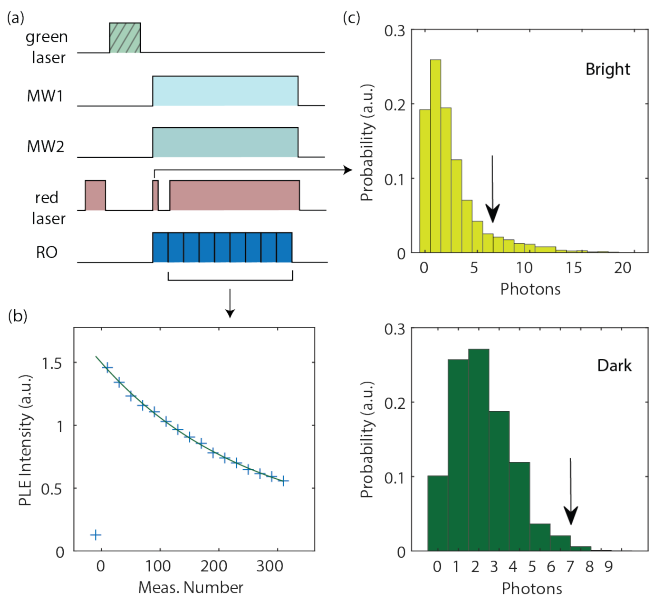

**Fig. S4: Multiple readout and post-selection.** (a) Experimental protocol. Upon charge initialization (via a red pulse and an optional green pulse), we apply simultaneous red and MW excitation during a prolonged time; we break out the readout time into smaller intervals (20 ms) and separately record the total number of photons in each time bin as we repeat the protocol multiple times. (b) $NV^-$ ionization curve as a function of the bin number; the total number of averages is 5000. (c) Photon count probability distributions as derived from the readout in an individual bin (the first one in the series after charge initialization, see schematics in (a)). Green or red illumination (respectively, upper and lower histograms) prepares the NV into "bright" or "dark" states; the arrows indicate the photon count threshold we use to distinguish between the two and serves as the criterion for post-selection. Note the short readout time (1 ms), which ensures minimal impact on the NV charge state. The red laser power is 50 nW.

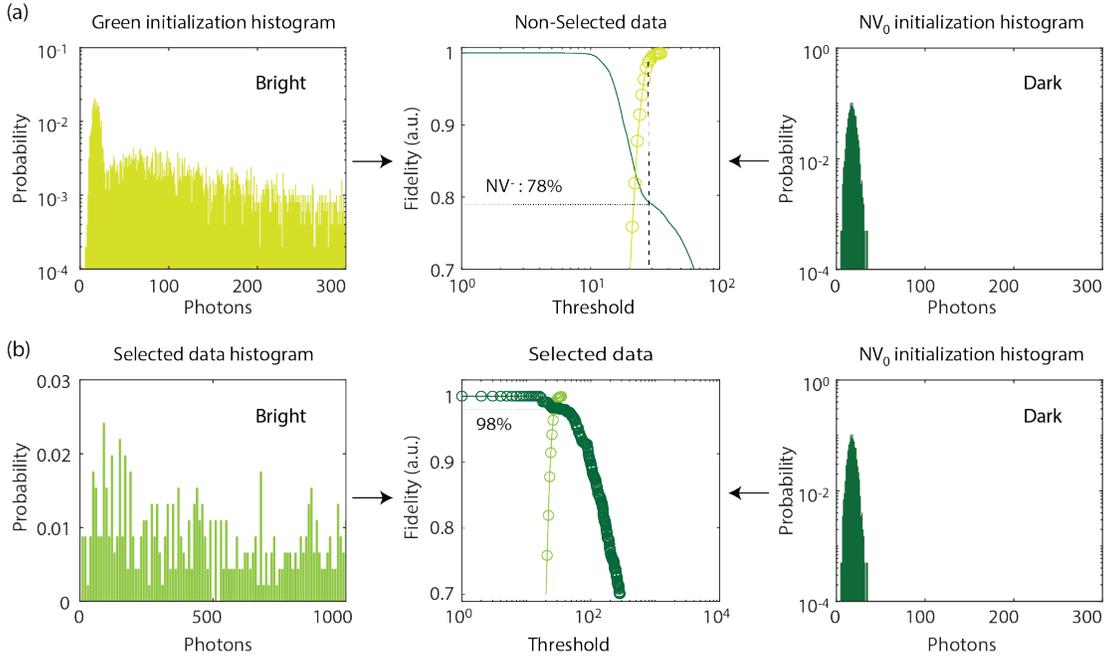

**Fig. S5: Single-shot charge readout fidelity.** (a) Calculated fidelity (center plot) and photon count probability distributions upon prolonged NV illumination with green or red (left- and right-hand side plots, respectively). Note that green excitation yields NV$^-$ (NV$^0$) with 78% (22%) probability. The red laser power is 50 nW and the readout time is 340 ms. (b) Same as in (a) but after post-selection using a protocol similar to that in Fig. S4a; with the conditions listed in (a), we attain an optimal charge readout fidelity of 98%. Dotted lines show the threshold and associated bright state read-out fidelity.

To address this problem, we make use of a short non-destructive read-out of the charge state to enhance the initialization fidelity. We then calculate the fidelity of a subsequent longer read-out (Fig. S4a). The key parameter here is the duration of the initialization read-out (1 ms for 50 nW), which we bring down to a minimum so as to avoid ionization. We confirm this condition through the experiments in Fig. S4b, where we break down the subsequent interrogation time into a collection of 20 ms readout bins; we find that on average hundreds of 1 ms read-outs are needed to produce appreciable NV$^-$ ionization. Figure S4c shows the probability distributions as obtained from a single readout bin under a 1 ms short laser pulse following green or red initialization (respectively producing "bright" and "dark" states as shown in the upper and lower histograms). Establishing a photon count threshold of 7 or more allows us to isolate events with NV$^-$ formation with 98% confidence, at the cost of eliminating 90% of the cases.

The ability to identify the initial NV charge state now allows us to determine the readout fidelity through a post-selection protocol where the probability distribution representing the "bright state" emerges exclusively from the subset of experiments where green illumination yields a negatively charged NV (i.e., experiments where the short initialization readout as implemented in Fig. S4 yields 7 photons or more). Figure S5 compares the readout fidelities obtained by standard means and through the post-selection protocol (respectively, data sets in the upper and lower rows); for simplicity, here we make the readout duration as long as possible with the understanding that shorter readout times will necessarily lower the fidelity. The bright (dark) state read-out fidelities are measured as the probability to measure a bright (dark) state given that the NV was initialized in the bright (dark) state. The threshold to distinguish between dark and bright state is taken to maximize both fidelities, using charge state post-selection for the bright state fidelity. In its absence (Fig. S5a), we find an optimal bright state readout fidelity of ~78%, consistent with reported NV$^-$ yield of green illumination at low temperatures[9]. On the other hand, the readout fidelity we extract from the post-selected data set reaches up to 98% for a red laser power of 50 nW. While spectral diffusion is compensated by power-broadening above 100 nW, it becomes a limitation for low power experiments. This is illustrated in the left histograms of Fig. S5, where the photon distribution is dominated by the shift between the laser and the NV line, resulting in a long tail corresponding to on-resonance events. Adjusting the post-selection protocol[6,9] or modulating the laser frequency to sweep the NV line would help increase the fidelity further. The results in Fig. 2c of the main text derive from similar experiments at varying red laser power. Note that because the charge readout time is made as long as possible, ionization ultimately takes place, hence making the process destructive. This said, non-destructive single-shot readout of individual NVs is still possible as the experiments in Fig. 3e of the main text demonstrate; we return to this point in Section 4 immediately below.

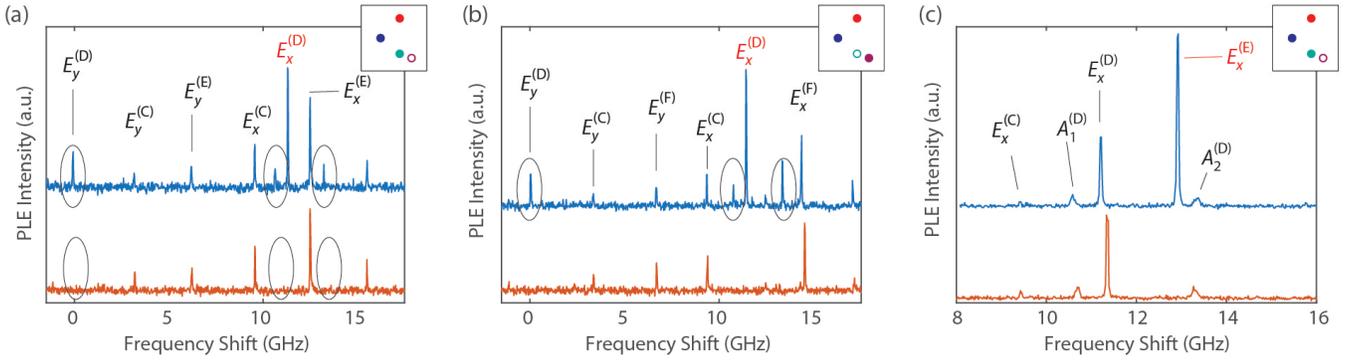

**Fig. S6: Charge control of an NV cluster.** (a) Optical spectroscopy of the four NV cluster in Fig. 3 of the main text for the case where all NVs are negatively charged except $NV_F$. Blue and red traces respectively show the spectra before and after selective ionization via laser excitation resonant with the transition denoted in red ($E_x^{(D)}$); ellipses enclose other transitions associated to the same NV. (b) Same as in (a) but for a different initial charge configuration where only $NV_E$ is neutral. (c) Same as in (a) but for selective ionization of $NV_E$. In (a) through (c), the inset in the upper right denotes the starting charge configuration using the notation in Fig. 3, namely, full (open) circles indicate negatively charged (neutral) NVs; red, blue, green, and brown circles respectively denote $NV_C$, $NV_D$, $NV_E$, and $NV_F$. All spectra have been displaced vertically for clarity.

**4. Charge control and temporal stability of NV clusters in Crystal A**

Although the preferential charge state of a trap ultimately emerges from the alignment of the trap ground state relative to the host crystal Fermi level, point defects such as the NV center can adopt metastable charge configurations often featuring virtually unlimited lifetimes. This is particularly the case in crystals of higher purity where spontaneous charge transfer processes in the dark become exponentially less likely as the average inter-defect distance increases. For example, charge tunneling from nitrogen donors should be negligible if the average inter-nitrogen distance exceeds 5-10 nm; with a nitrogen content of only a few ppb, this is certainly the case for Crystal A.

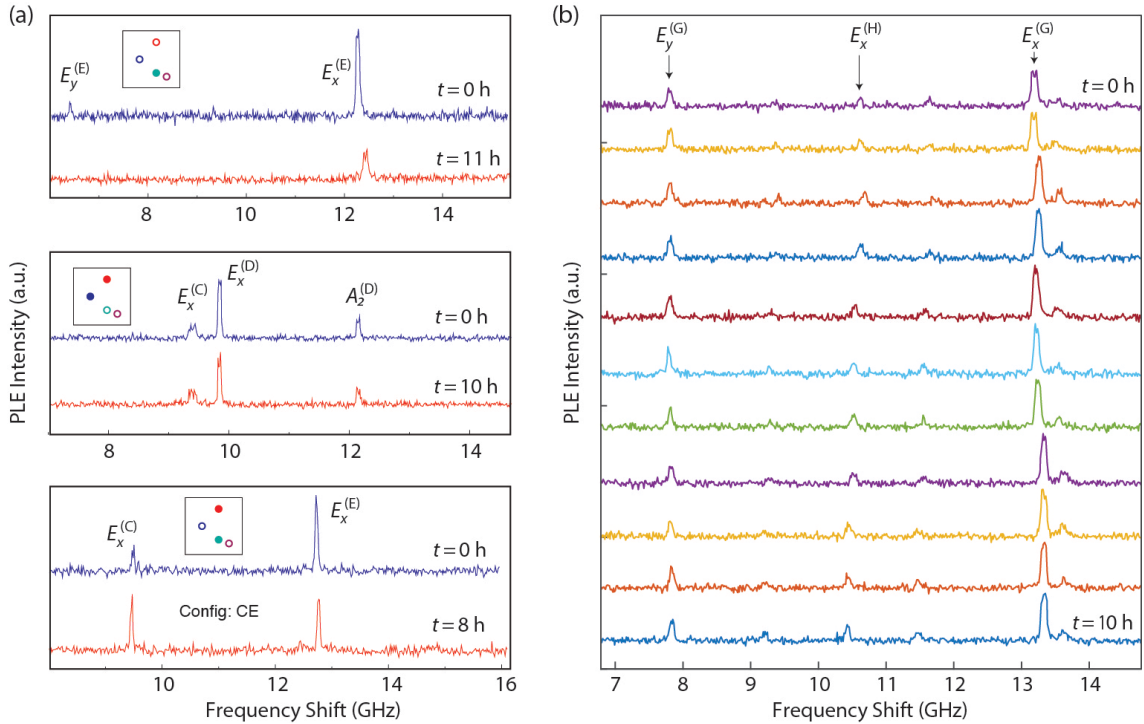

**Fig. S7: Temporal stability of NV clusters.** (a) PLE spectroscopy of the four-NV cluster in Fig. 3 of the main text. The cluster schematic indicating the starting charge state configuration (inset) follows the notation in Fig. 3 of the main text (see also Fig. S6). The upper and lower trace (here displaced vertically for clarity) respectively show the spectra immediately after charge initialization and after varying time intervals in the dark. (b) Repetitive readout of a two-NV cluster (here denoted $NV_G$ and $NV_H$) over a 10-hour window; the interval between successive readouts is 1 hour and the observation time grows from top to bottom.

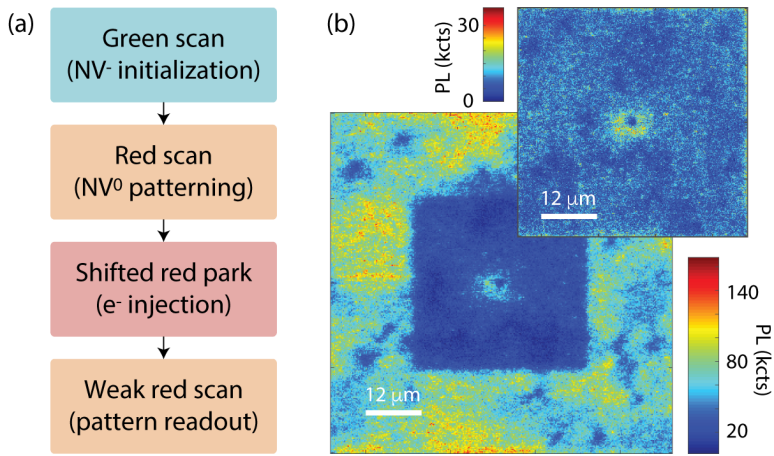

**Fig. S8: Capture of itinerant electrons.** (a) Experimental protocol. Following green initialization into NV⁻, we pattern an NV⁰-rich square using red illumination at 470.48 THz; we then shift the laser frequency by 20 GHz and park the beam at the square center. Finally, we bring back the laser frequency to its original value and take a scanning fluorescence image using the protocol in Fig. 1b of the main text. (b) Experimental result following the sequence in (a). The upper right insert is a zoomed reconstruction of the charge pattern at the center of the imaged plane for a red laser park time of 120 s. The bright ring around the point of illumination indicates recombination of NV⁰ into NV⁻ via electron capture. The temperature in these experiments is 7.5 K.

The spatial distribution of nitrogen, however, can be highly heterogeneous, thus raising questions on the ability to reliably control the charge state of a cluster and maintain its charge configuration over a long time interval. Fig. S6 summarizes some observations for varying starting charge configurations and/or selective ionization of different NVs in the cluster of Fig. 3 in the main text; combined with the results in Fig. 3e, demonstrate we demonstrate virtually full control of the cluster charge state. Fig. S7 shows additional observations primarily aimed at gauging the temporal stability of a given charge state both for the four-NV cluster and for a new two-NV system (here denoted $NV_G$ and $NV_H$). In all cases, we find the starting charge configuration remains unchanged over several-hour-long periods, hence indicating spontaneous NV ionization (or recombination) in these systems is rare; in passing, our experiments also demonstrate repetitive non-destructive charge readout, here made possible by resorting to sufficiently low laser powers (1 nW).

Unfortunately, extending the experiments in Fig. S7 over longer time intervals is presently difficult because unintended system drifts complicate our ability to relocate the cluster for inspection (note that using green or strong red laser scans may inadvertently alter the cluster charge state). Future work can address this problem, e.g., with the aid of reference grids patterned on the diamond surface. Lastly, it is worth mentioning that while the present results rely on a probabilistic charge initialization into NV⁻, NV⁰ is known to possess narrow zero-phonon transitions at 575 nm[10], suggesting that the use of narrow-band resonant excitation could possibly be exploited to deterministically induce NV⁰ recombination into NV⁻.

## 5. Multiple consecutive readouts and data multiplexing in ensembles

We saw in Sections 3 and 4 how successive readouts of a single negatively charged NV necessarily increase the chance of ionization and correspondingly lead to a decay of the charge state readout fidelity. Given the two-photon nature of the ionization process, minimal impact on the NV⁻ charge state requires lower laser intensities, which translate into increasingly longer readout times. While a similar tradeoff applies to ensemble measurements, the complexity of higher defect concentrations brings to the fore other phenomena ignored in the limit of small color center sets.

The first one we consider herein relates to unintended diffusion and recapture of photo-generated carriers. Prior work has shown how optical excitation of diamond can lead to color center ionization and recombination, respectively resulting in the injection and diffusion of itinerant electrons and holes[3,4,7]. For the present experiments, red illumination can produce ionization of NV⁻ and of neutral nitrogen impurities (the so-called P1 centers) hence resulting in the preferential photo-generation of diffusing electrons. In principle, capture of some of these carriers by NV⁰ can alter a pre-existing charge pattern and thus induce information loss.

We investigate this possibility through the experimental protocol in Fig. S8a, where we monitor the impact of a red laser park on the charge state of proximal NV⁰. As shown in the images of Fig. S8b, the bright ring around the point of illumination signals electron capture by NV⁰; this process — absent under ambient conditions[11] — amounts to a larger electron capture cross section at cryogenic temperatures, which, in turn, suggests the stable formation of weakly bound orbits[12]. Nonetheless, the long park times required to make the process observable (of the order of minutes) indicates that the impact of electron capture on pre-existing NV charge patterns should be low. We confirm this notion in Figs.

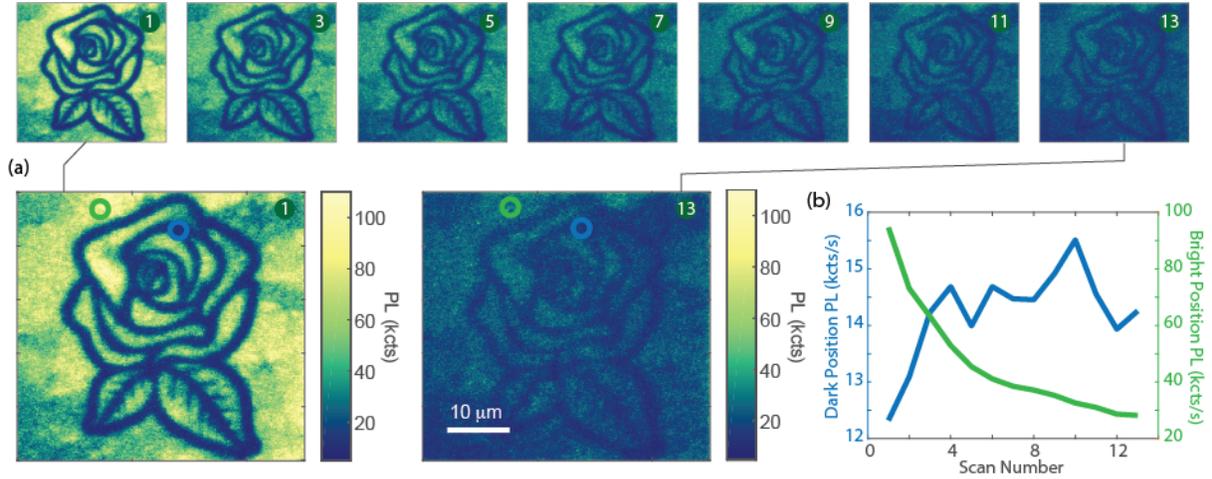

**Fig. S9: The impact of multiple readout.** (a) Images of the same charge pattern upon consecutive readouts. The number in the upper right corner indicates the scan number. (b) Fluorescence as a function of the scan number at two different locations originally imprinted as "bright" and "dark" (respectively, orange and blue circles in the images in (a)). Successive readouts lead to an exponential decay of the bright zone due to progressive NV$^-$ ionization. Upon a few readouts, we also observe a mild increase in the fluorescence of the dark zone, possibly due to electron diffusion and capture; this process comes to a stop once the brightness of adjacent areas is sufficiently low.

S9a and S9b, where we probe the fluorescence of bright and dark sections of a pattern as we implement consecutive readouts. Despite an early growth, we find that the fluorescence from dark areas subsequently stabilizes and remains comparatively low (less than 1/6 of the maximum). As shown in Fig. S10, consecutive readouts have a similarly minor effect on dark areas even in the limit of high data multiplexing. This said, additional work will be needed to assess the impact of diffusing holes on bright sections of a pattern, especially if resonant laser excitation near 575 nm allows one to selectively recombine NV$^0$ into NV$^-$. On a related note, we mention that in Figs S9 and S10 the unintended NV$^-$ ionization during subsequent readouts can be mitigated by reducing the readout power and increasing the readout time; as in Fig. 3 of the main text, however, a similar tradeoff between fidelity, laser power, and readout time applies.

Another consideration — particularly important for large NV ensembles — relates to the crosstalk between data sets in nominally separate frequency windows. Since each individual NV features multiple optical resonances, transitions from distinct color centers necessarily overlap as the PLE spectrum gradually becomes a smooth continuum. Therefore, correlations between charge patterns carved out at different frequencies are to be expected, even though the level of crosstalk is a priori unclear; also uncertain is whether these correlations are spectrally (or spatially) confined.

Fig. S11 summarizes the results from an experiment designed to tackle some of these questions: Here we use resonant light to write a simple NV charge pattern, which we subsequently probe through successive imaging scans at gradually shifted resonances. We find virtually no impact on NVs separated by frequency differences in excess of ~6 GHz, a value

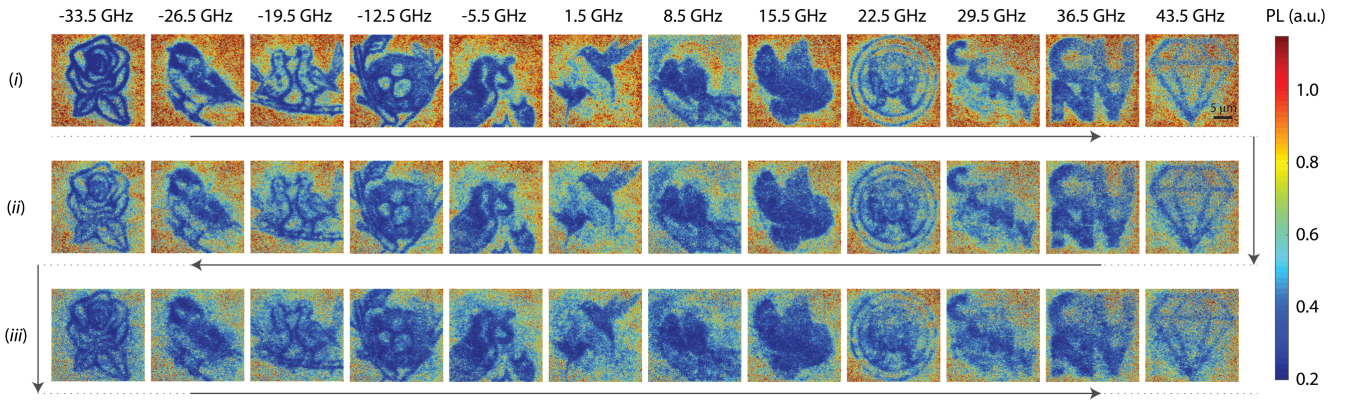

**Fig. S10: Multiple readout and data multiplexing.** Reproducing the data set in Fig. 4 of the main text, we carry out three successive readouts of all frequency-encoded layers (rows (*i*) through (*iii*)) in alternating order, as indicated by the horizontal arrows. All experimental conditions reproduce those in Fig. 4 of the main text.

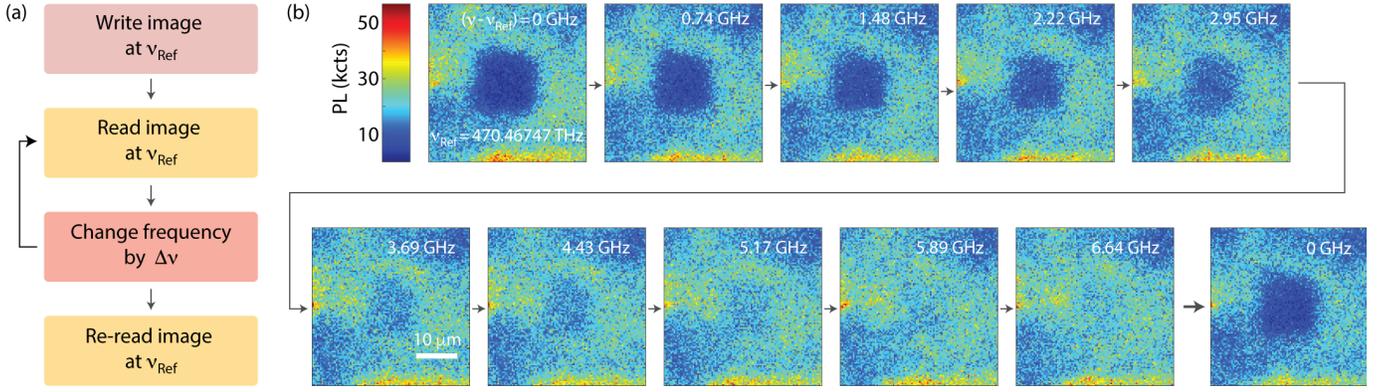

**Fig. S11: Multiplexing density.** (a) To bring the cross talk between consecutive images in a multiplexed stack to a minimum, we set the red laser frequency to $\nu_{\text{Ref}}$ so as carve out a pattern of neutral NVs we can use as a reference; we subsequently probe the diamond same plane at gradually shifted frequencies. (b) Experimental implementation of the protocol in (a). For the present conditions, a 6.6 GHz shift is sufficient to make the cross talk virtually negligible; from re-reading the charge map at $\nu_{\text{Ref}}$ (last image in the series), we confirm the original pattern remains virtually unchanged.

consistent with the spectral span observed in individual NVs under low crystal strain (Fig. S1).

To make our assessment quantitative, we define a crosstalk matrix as[13]

$$\rho_{ij} = \frac{\text{Cov}(T_i, R_j)}{\sqrt{\text{Var}(T_i)\text{Var}(R_j)}} - \Delta\rho_{ij}, \quad (S1)$$

where $T_i$ and $R_j$ respectively denote each of the $N$ target and imprinted images in a multiplexed plane, and Cov (Var) indicates co-variance (variance) calculated by summing over all pixels in each image. Using $\delta_{i,j}$ to indicate Kronecker's delta, $\Delta\rho_{ij} = \text{Cov}(T_i, T_j)/\sqrt{\text{Var}(T_i)\text{Var}(T_j)} - \delta_{i,j}$ represents an off-diagonal correction term that takes into account unintended correlations in the chosen target data set. The average across the matrix diagonal is a measure of the (combined) writing and reading fidelity, while off-diagonal elements capture cross-correlations between nominally independent groups of data. Applying the above formula to the images in Fig. S11b, we find that $\rho_{1,10}$ — the crosstalk coefficient between images 1 and 10 in the series — amounts to 3%, hence establishing a characteristic "frequency correlation span" that we subsequently build upon for efficient data multiplexing.

Extending these calculations to the data set in Fig. 4 of the main text is complicated in that some of the target images use a gray scale, thus creating ambiguities in the resulting fidelities. We circumvent this problem by transforming all images (both in the target and imprinted sets) to a "digital" bright/dark scale using a threshold of 0.5 (Fig. S12a). The matrix in Fig. S12b reproduces the results: Given the 7 GHz separation between successive frequency layers in the set, we find that inter-layer crosstalk remains moderate or low.

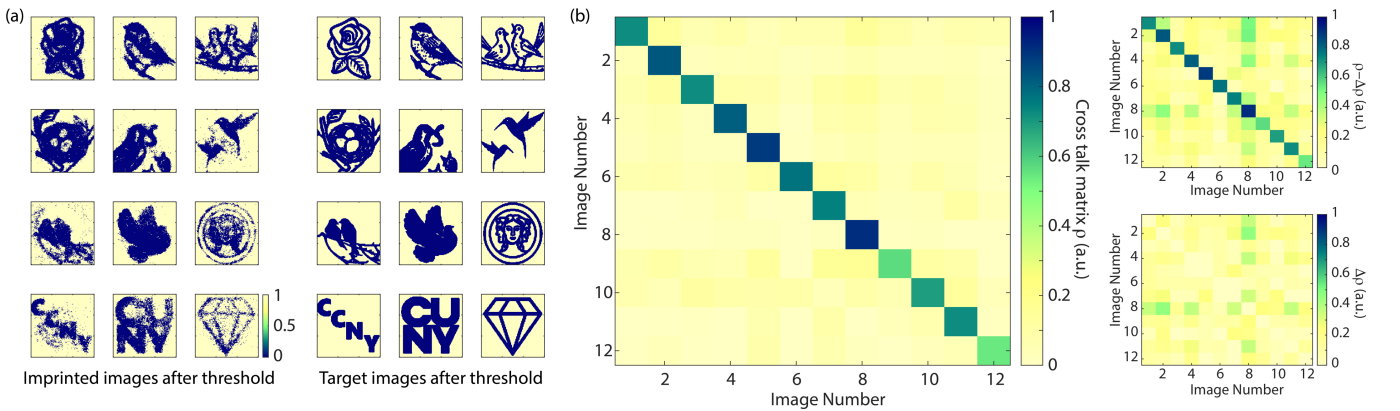

**Fig. S12: Gauging the impact of cross-correlation.** (a) Imprinted and target images as derived from the set in Fig. 4 of the main text (left and right set, respectively) after digitization using a relative amplitude threshold of 0.5. (b) Cross-correlation matrix $\rho$ for the set in (a). The average correlation (writing and reading fidelity) is below 0.05 (above 0.75). The upper and lower right inserts separately show the correlation matrix with no correction and the correction matrix, respectively; comparison with the main plot shows that $\Delta\rho$ is relatively small.

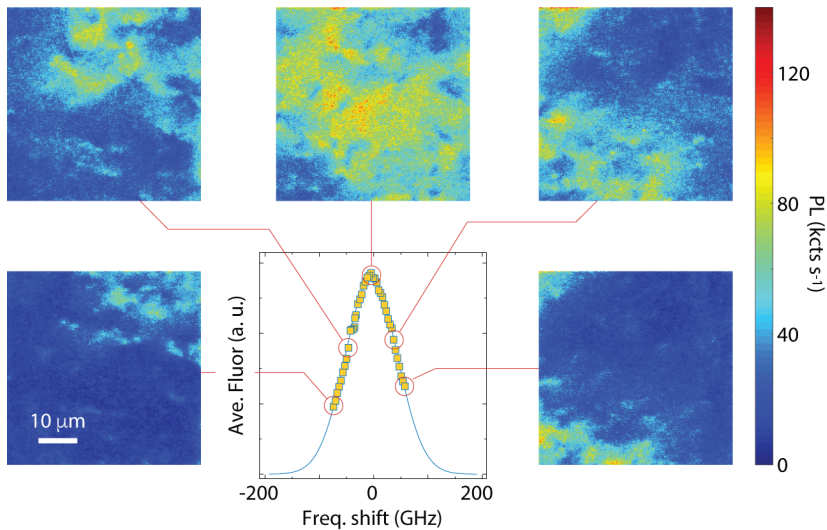

**Fig. S13: Connecting spectral and spatial NV heterogeneity.** (Main) NV distribution as a function of the red laser detuning relative to 470.48039 THz; each point in the spectrum derives from averaging the NV$^-$ fluorescence over a 50×50 μm$^2$ area. (Side inserts) NV$^-$ fluorescence images at select optical frequencies (circles in the main spectrum). Despite the heterogeneity (slightly higher here than in other areas of the crystal), we find that an increase in the laser frequency leads to the gradual displacement of bright sections from the lower-left to the upper-right corner. We tentatively attribute this long-range spatial correlation to the impact of crystal strain. All experimental conditions similar to those in Fig. 4 of the main text.

Applications to data storage aside, the results in Fig. S10 provide some valuable clues on the statistics governing the NV transition frequencies, a piece of information that will likely prove helpful when connecting microscopic models of heterogeneity to the Hamiltonian governing the NV$^-$ $^3E$ fine structure[14]. For example, the fact that the frequency correlation span is much smaller than the overall inhomogeneous linewidth should translate into corresponding differences in the statistical distributions assigned to the longitudinal and transverse components of the NV$^-$ electric dipole moment. Note that since the coupling terms connecting these dipole components to electric or strain fields are formally equivalent, the observations in Fig. S10 alone are insufficient to single out one contribution or the other.

One possibility to discriminate between these two sources is to investigate their spatial dependence. For example, one would expect strain fields to change more gradually across the crystal and hence lead to long-range spatial correlations presumably absent when considering the electric environment. The observations in Fig. S13 seem to confirm this notion: Here we compare several images obtained for a section of Crystal B under 637-nm excitation, and immediately find that resonant NVs preferentially move from the upper-right to the lower-left corner of the focal plane with increasing red laser frequencies. Since frequency multiplexing as shown herein rests on NV ensembles within the same focal volume, these results suggest electric fields emanating from carriers in deep traps would be responsible for the inhomogeneity. We caution, however, that it would be premature to rule out crystal strain as an equally important contributor, particularly considering the major impact it can have on the NV optical response (Fig. S1).

## 6. Spatial resolution and interplay between adjacent pixels

In the absence of wavelength multiplexing, the data density is related to the spatial resolution of the write-read protocol, which is limited by optical diffraction in the optical *xy*-plane and the beam divergence in the optical *z*-axis. Here, we perform an estimation of those limits in our setup (NA=0.75) by patterning an array of dark dots at a single frequency, as shown in Fig. S14a. In the optical plane, fitting 42 dots with a Gaussian distribution, we find an average

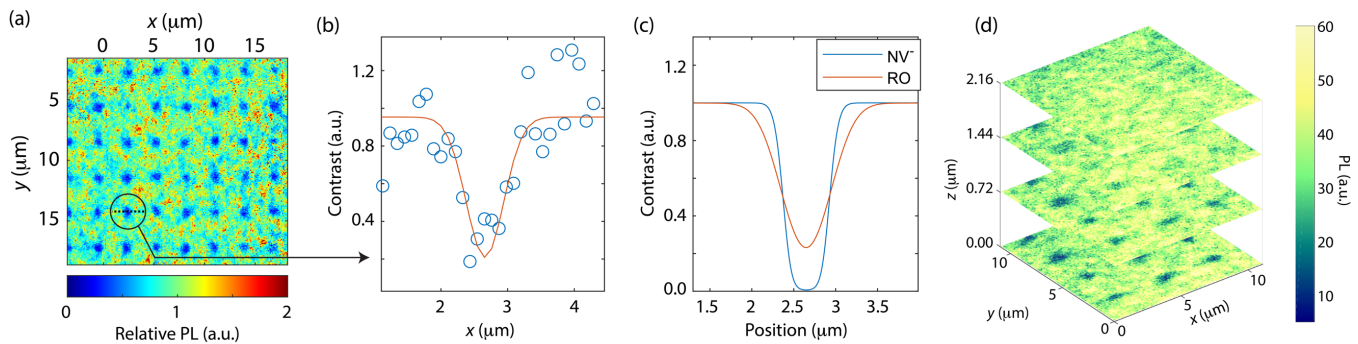

**Fig. S14: Spatial resolution.** (a) Array of dark dots imprinted on a bright background. (b) Horizontal linecut of the 2D gaussian fit yielding a FWHM of 670 nm for this dot. (c) Expected profile of the dot: NV$^-$ ionization probability and optical readout (RO). (d) PL slices at different depth of a dot array patterned at *z* = 0. Write (read) parameters are 30 μW for 5 ms (200 nW for 5 ms).

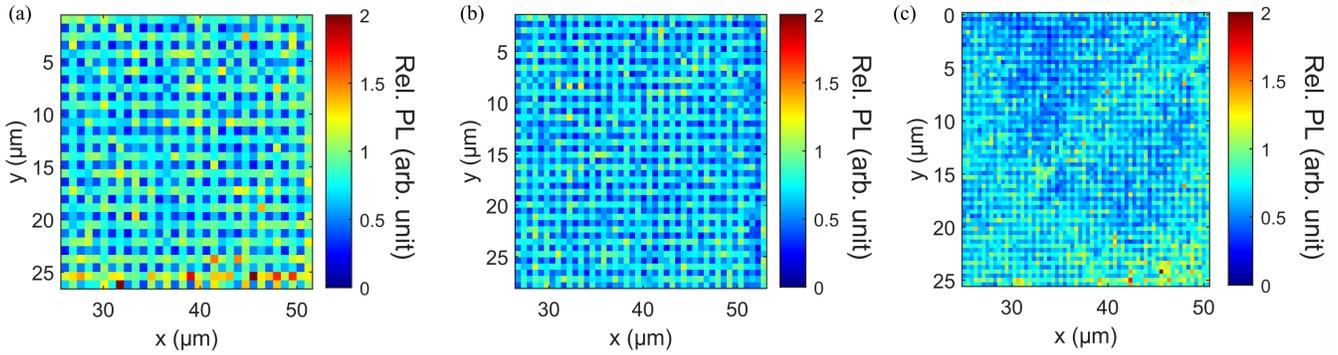

**Fig. S15: Interplay between adjacent pixels.** All figures show arrays of dark dots imprinted on a bright background for varying inter-dots distance: (a) 800 nm, (b) 600 nm, and (c) 400 nm, yielding fidelities of 0.99, 0.97, and 0.75 respectively. Write (read) parameters used are 20 µW for 5 ms (200 nW for 50 ms).

FWHM (contrast) of 890 nm (0.64), going as low as 670 nm (0.79) for specific sharper dots (Fig. 14b). Those values, well-above the laser spot-size of the system (467 nm for the resonant laser) arise from a convolution of the writing and reading protocols. Figure 14c shows the expected profiles along the $x$-axis for the $NV^-$ ionization probability and photoluminescence of one dot. We assume an ionization rate proportional to the intensity square and take a dwell time five times as long as the inverse ionization rate at the maximum intensity. We find a FWHM of 684 nm for a contrast of 0.77, in excellent agreement with values measured from sharper dot imprints. Along the optical axis, Fig. 14d shows that patterns are indistinguishable from the background 2.16 µm away from the imprinted plane, giving a full width along the $z$-axis of roughly 2.30 µm.

To estimate the maximum data density in the optical plane, we measure the read-write fidelities in arrays of dots separated by diminishing distances (800, 600 and 400 nm). The PL read-out from those arrays is depicted in Fig. 15: Here, we combine the read-out fidelities of dark pixels (center of the imprinted dots) and bright pixels (in-between dots, along the vertical direction in this experiment). Fidelities are obtained using a threshold for the relative PL (around 0.55, optimized for average fidelity of bright and dark pixels). We find average fidelities of 0.99, 0.97 and 0.75 for 800, 600 and 400 nm distances respectively, showing minimal interplay for distances above 600 nm. Interestingly, we noticed that the fidelity is impacted by inhomogeneities in the NV canvas (e.g., bright diagonal strip in Fig. S15c), which could stem from varying NV density or varying strain and provides a clear pathway for improvement. By performing a 12× wavelength multiplexing as in the main text, we can therefore bring the bit density from $(1 \text{ bit})/(600 \text{ nm})^2$ to $(12 \text{ bit})/(600 \text{ nm})^2 = (1 \text{ bit})/(170 \text{ nm})^2$, well-past the diffraction limit.

One limitation in the storage density arises from the finite NV content, particularly if the spatial distribution of color centers on the plane is to remain reasonably uniform. While greater NV concentrations can potentially mitigate the problem, a critical threshold is to be expected as metastable charge states gradually succumb to Fermi statistics. From prior observations[15], we tentatively set the upper bound for NV charge metastability below 100 ppb, though we caution this value can vary largely depending on the concentration of co-existing charge traps, most notably vacancies as well as nitrogen and silicon impurities. For example, while charge states are long-lived in Crystal B at cryogenic temperatures, we observe some data loss after a four-day thermal cycle from 4 K to room temperature and back (see immediately below). Interestingly, the NV charge state has been seen to remain stable in diamonds with similar NV content (but different composition) over periods exceeding a week at room temperature[15], hence indicating cryogenic temperatures may ultimately not be required for data storage.

## 7. Charge stability in NV ensembles; the impact of a thermal cycle

A natural question is whether the multiplexed patterns we create herein under cryogenic conditions remain unchanged at higher temperatures. We explore the impact of temperature on the charge state stability of NVs in Fig. S16 where we subject the ensemble-hosting diamond used in Fig. 4 of the main text (Crystal B) to a thermal cycle from 7 K to room temperature and back. The schematics in Fig. 16a lay out our protocol: We use spectrally selective NV ionization to encode (and readout) four different data sets multiplexed into the same optical plane at 7 K. After warming up the system to ~300 K, we bring the system back to cryogenic conditions for wavelength-selective readout; completion of the thermal cycle requires an interval of four days. Fig. S16b shows the results: Comparison of the charge patterns before and after the thermal cycle reveals a partial loss of contrast, in this case predominantly produced by $NV^0$ recombination into $NV^-$. We quantify this loss in Figs. S16c and S16d where we calculate the correlation matrix upon digitizing the image set: From the average value along the matrix diagonal, we find a drop from 0.84 to 0.75; the crosstalk between images in this case amounts to about 1%.

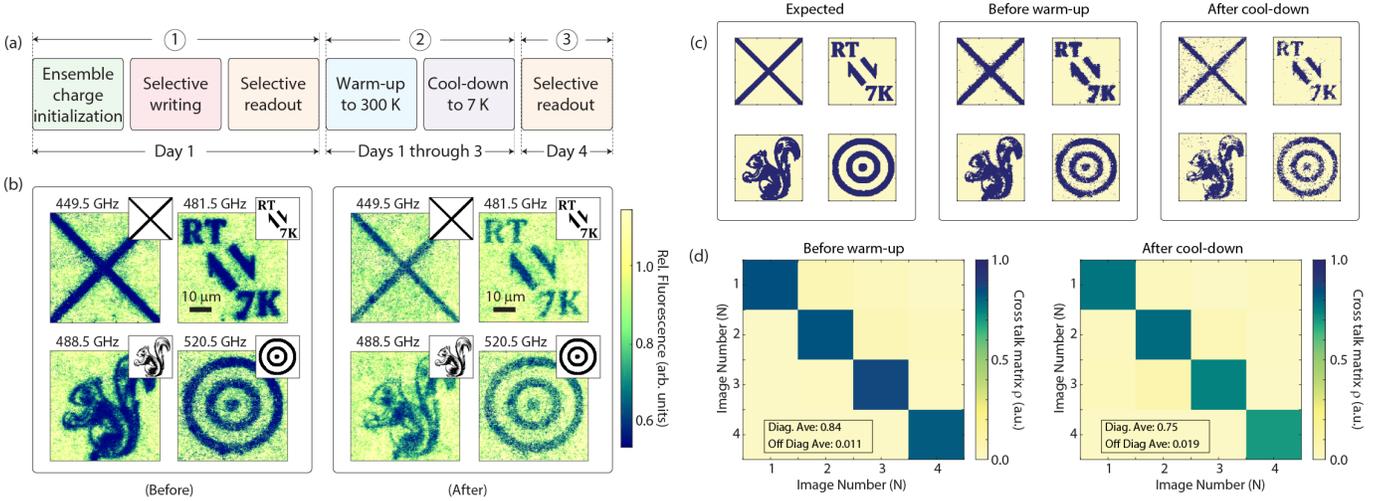

**Fig. S16: The impact of thermal cycling.** (a) Experimental protocol; upon data writing and readout (step ①), the sample undergoes a thermal cycle comprising warm-up and cool-down ramps from 7 K to room temperature and back (step ②), followed by a second readout round (step ③). (b) (Left) We follow the protocol in Fig. 3c to write and read a multiplexed data set in a diamond plane at 7 K. (Right) Retrieved data set after the thermal cycle. Despite the reduced contrast, all patterns are still clearly distinguishable. All experimental conditions as in Fig. 3. (c) Target (left set) and imprinted images upon digitization before the warm-up and after the cool-down (middle and right sets, respectively). (d) Calculated correlation matrices before and after the thermal cycle (left and right, respectively). The average write/read fidelity averages 0.84 (0.75) before (after) the cycle; the average cross-correlation in both cases is ~0.01.

We tentatively attribute this loss of fidelity to thermally activated electron transfer from neighboring donors (most likely substitutional nitrogen), which could perhaps be inhibited in diamonds hosting a sufficiently large concentration of electron traps. Indeed, previous observations in a crystal with similar nitrogen concentration (~1 ppm) but also hosting SiVs (at ~100 ppb and known to preferentially take SiV$^-$ and SiV$^{2-}$ charge states) indicate no changes in previously NV crafted charge patterns over a time span of a week at room temperature[15].

## 8. Outlook: Opportunities and challenges

Although an in-depth comparison between the present approach and other non-volatile data storage technologies is premature, the use of color centers as a memory platform — either in the form of NVs or related point defects — promises some intriguing opportunities worth exploring in further detail. For example, as in holographic data storage, the entire three-dimensional crystal serves as the recording medium and data writing and reading can be potentially carried out in parallel by resorting to a wide-field geometry, without the need for mechanical motion (by contrast, the case for blue-ray disks or magnetic hard-drives). Further, since related physical principles (involving the photo-generation and redistribution of trapped carriers) govern pattern formation in optical holography and color center charge control, one could anticipate a similarly long storage lifetime, though sub-diffraction access arguably makes the storage density possible through the present approach comparatively higher. Something similar can be said when compared to other emerging optical storage strategies such as super-resolution lithography[16], a "write once, read many" (WORM) approach limited to surface arrays.

Along related lines, other non-optical memory platforms feature higher storage densities, but geometric and practical constraints raise various problems difficult to overcome. One illustration is electronic quantum holography, a technique reaching storage densities of up to 20 bits/nm$^2$ but limited to surfaces and difficult to scale up given the comparatively slow write times; further, the conditions required for scanning tunneling microscopy (such as high vacuum) complicate practical implementations (a limitation also applicable to color center charge control given the present need for cryogenic temperatures). Similarly, "DNA digital data storage" — the process of encoding and decoding binary data to and from synthesized strands of DNA — can attain storage densities of ~200 petabits/gram though the high-cost per run along with characteristically long write and readout times do not compare well with the approach herein.

While extensive additional work is needed to better assess the viability of charge state control as a practical data storage technology, our work provides a starting point by (*i*) demonstrating that the NV charge state can be controlled individually in groups of NVs separated by sub-diffraction distances (Fig. 2 and 3 in the main text), and (*ii*) showing it is possible to separately address NV subsets in a diffraction limited volume even if selectively addressing individual

NVs becomes impractical (Figs. 4 in the main text along with Figs. S14 and S15). Note that despite all related prior work, neither result is a priori guaranteed as one may argue, e.g., that recapture of diffusing carriers by NVs separated by sub-diffraction distances can easily bring down the writing fidelity to impractically low values. Further, it is not immediately clear that one can reduce the crosstalk between spatially overlapping, charge encoded NVs by sufficiently detuning the excitation laser (as a matter of fact, one would expect the opposite to be true).

Looking forward, an area key to addressing individual emitters in denser ensembles is sample engineering: Studying the response of crystals with variable NV content is central to better defining the boundaries where charge states become unstable; equally important is attaining control of the sample composition and better characterizing the impact on NV charge stability from other point defects in the diamond lattice. In the same vein, future work should attain a better understanding of the nature and physical mechanisms underlying the spectral heterogeneity we observe in the NV optical transitions. Specific goals in these studies should be (*i*) discriminate between contributions from local electric fields and crystal strain, (*ii*) establish the fundamental rules governing the interplay between the average shift and frequency spread in the set of NV optical transitions, and (*iii*) better understand the impact of increasing spectral heterogeneity — both, longitudinal and transverse — on NV brightness.